\RequirePackage[2020-02-02]{latexrelease}
\documentclass[twocolumn,preprintnumbers,amsmath,amssymb,superscriptaddress]{revtex4}
\usepackage{graphicx}
\usepackage{dcolumn}
\usepackage{bm}
\usepackage[T1]{fontenc}
\usepackage{xcolor}

\begin{document}

\title{Wave-packet dynamics in monolayer graphene with periodic scattering potentials}

\author{M.M. Suleimanov}
\affiliation{Institute of Materials Science, Uzbekistan Academy of Sciences, 2-B Chingiz Aitmatov St., Tashkent 100084, Uzbekistan}
\affiliation{Tashkent International University of Education, 31 Imam Bukhariy St., Tashkent 100207, Uzbekistan}
\author{M.U. Nosirov}
\affiliation{Institute of Materials Science, Uzbekistan Academy of Sciences, 2-B Chingiz Aitmatov St., Tashkent 100084, Uzbekistan}
\author{H.T. Yusupov}
\affiliation{Department of Exact Sciences, Kimyo International University in Tashkent, 156 Shota Rustaveli St., Tashkent 100121, Uzbekistan}
\author{A. Chaves}
\affiliation{Departamento de F\'isica, Universidade Federal do Cear\'a, Caixa Postal 6030, Campus do Pici, 60455-900 Fortaleza, Cear\'a, Brazil}
\author{G.R. Berdiyorov}
\affiliation{Qatar Environment and Energy Research Institute, Hamad Bin Khalifa University, Doha, Qatar}
\author{Kh. Rakhimov}
\email{kh.rakhimov@gmail.com}
\affiliation{Central Asian University, 264 Milliy bog St., Tashkent 111221, Uzbekistan}
\affiliation{Institute of Materials Science, Uzbekistan Academy of Sciences, 2-B Chingiz Aitmatov St., Tashkent 100084, Uzbekistan}
\date{\today}

\begin{abstract}

We use the Dirac continuum model to study the propagation of electronic wave packets in monolayer graphene in the presence of periodically arranged circular potential steps. The time propagation of the wave packets is calculated using the split-operator method for different sizes, heights, and separations of the barriers. We found that, despite the pronounced Klein tunneling effect in graphene, the presence of a lattice of defects significantly impacts the propagation properties of the wave packets. For example, depending on the height and size of the incident wave packet, the transmission probability can decrease by more than 30\%. The alteration of the polarity of the potential barriers also contributes to the transmission probabilities of the wave packets in graphene. The results obtained here provide valuable insights into the fundamental understanding of charge carrier dynamics in graphene-based nanodevices.

\end{abstract}
\maketitle

\section{Introduction}

\begin{figure}[t]
\centering
\includegraphics[width=\linewidth]{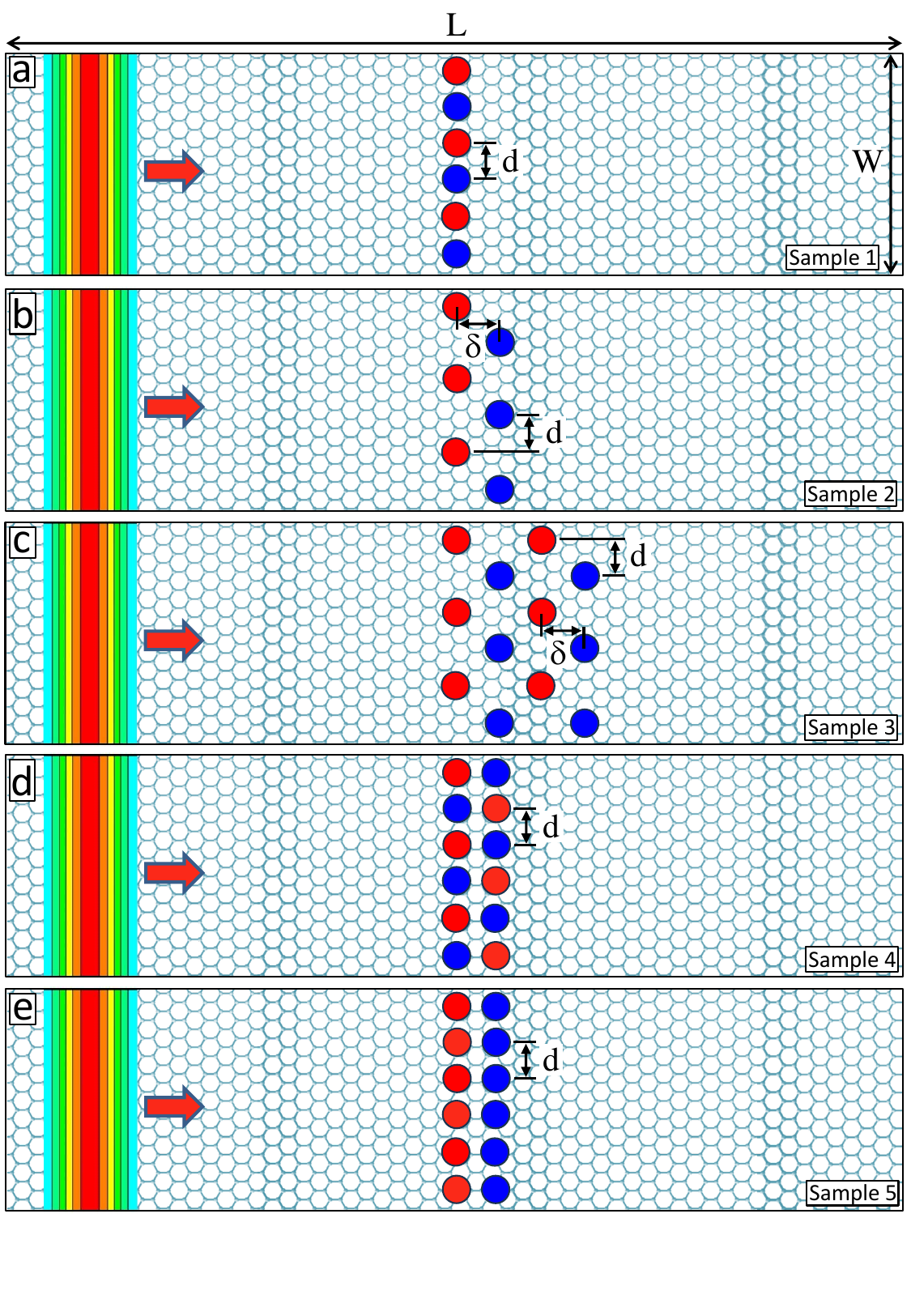}
\vspace{-1.cm}
\caption{\label{fig1} Schematic illustration of wave packet propagation in graphene along the forward direction (indicated by an arrow) in the presence of periodically arranged circular potentials. The red and blue dots represent regions with positive and negative potentials, respectively. Five configurations of dots are considered, labelled Sample 1-5.}
\end{figure}

Graphene is among the most promising two-dimensional (2D) materials for nanoelectronics applications, mainly due to its mechanical and thermal stability, as well as its exceptionally high carrier mobility and flux \cite{Novoselov2004,Geim2007,Geim_Science_2009,Castro2009,Novoselov2012}. However, chemically synthesized graphene frequently displays various structural irregularities, such as grain boundaries, vacancies, dislocations, impurity atoms, and defects arising from changes in carbon hybridization \cite{Gass2008,Meyer2008,Eckmann2012}. Consequently, these defects significantly affect graphene's structural stability, electronic and transport properties, optical characteristics, and other observable features, as even slight modifications to the honeycomb arrangement of carbon atoms can drastically impact these properties \cite{Hashimoto2004,Banhart2011,Telling2007}. Therefore, understanding the dynamics of charge carriers in the presence of structural defects is of practical importance for the development of graphene-based nanoelectronic components.

In this work, we use numerical simulations to investigate the propagation of wave packets representing low-energy electrons in graphene in the presence of a periodic lattice of scalar electrostatic potentials. These wave packets can scatter off such defects, ultimately influencing the material's conductance \cite{lin16,jia19,aps16,geza,chen,yuq,qiu}. Such scattering mechanisms have also been explored in different contexts, such as the formation of ballistic geometric diodes in graphene, where directional transport arises from geometry-induced asymmetry \cite{golib,jap25}. Extensive research has been conducted on wave packet propagation within the Dirac-Weyl Hamiltonian in pristine graphene, graphene with various dopants \cite{aps16,geza}, and under uniform external magnetic fields \cite{rusin1,rusin2,maks,rome}. However, theoretical studies on wave packet propagation through potential barriers remain limited \cite{nit99,pru09,khamdam,bal03,anker,kren,wat17,wat18}. In this study, we model a periodic lattice of circular charged regions (electron-rich and hole-rich puddles) that can form in graphene due to inherent disorder \cite{Martin2008,Schubert2012}. We implement the Dirac continuum model to understand the effect of such disorder on the transmission of electrons in graphene. This method was successfully applied recently to study wave packet propagation in graphene in the presence of randomly distributed electric potentials \cite{random, book13}. We found that, despite overall charge neutrality, the presence of such a periodic lattice significantly affects wave packet propagation in graphene. For instance, depending on the parameters of the wave packet, such as its energy, the transmission probability ($P$) can be reduced by up to 30\%.

\section{Model system and theoretical framework}

Figure \ref{fig1} illustrates our model systems, which comprise a graphene monolayer with lateral dimensions $L = 1024$ nm and $W = 128$ nm, featuring a periodic arrangement of scalar electrostatic potential dots. Each circular dot with radius $R$ mimics the presence of electron and hole puddles in real graphene samples, where such puddles appear with equal probability for both electrons and holes. To model this, we use scattering centers alternating between positive ($+V_0$) and negative ($-V_0$) potentials, representing regions with locally higher hole (positive potential) or electron (negative potential) densities, while maintaining a net zero average potential across the entire scattering region. The separation of the dots in the direction perpendicular to the wave packet propagation is fixed at $d = 21.3$ nm for all samples considered. For samples 2 and 3, the barriers are separated by a distance of $\delta = 6R$ along the direction of wave packet propagation, while for samples 4 and 5, this separation is $\delta = 3R$.

The electron wave packet is modeled as a Gaussian wave front characterized by energy $E$ and width $a_x$. The initial wave function is thus a wave front that is constant the $y$-direction, has a width $a_x$ along the $x$-direction, and is multiplied by a pseudo-spinor (1 ~ $i$)$^T$ and by a plane wave
\begin{equation}
\Psi(x,y,0) = N\left(\begin{array}{c}
1\\
i
\end{array}\right)e^{ikx - \frac{x^2}{2 a_x^2}}
\end{equation}
where $N$ is a normalization constant and $k$ is the wave vector, which, for low energy electrons in monolayer graphene, is related to the wave packet energy $E$ by $k = E \big/ \hbar v_F$, where $v_F$ is the Fermi velocity. The choice of the pseudo-spinor is made so that $\langle \sigma_x \rangle = 1$ and $\langle \sigma_y \rangle = \langle \sigma_z \rangle = 0$, which guarantees a wave packet propagation along the $x$-direction~\cite{chaves}. Moreover, we choose to work with a wave front, instead of a circularly symmetric wave packet, in order to avoid \textit{zitterbewegung} (a trembling motion of the wave packet)~\cite{chaves, rusin} along the $y$-direction. All the results in the remainder of this paper are obtained for a wave packet with energy $E = 100$ meV and widths $a_x = 100,~150$ and 200~nm.

The propagation of the wave packet is performed by applying the time-evolution operator on the initial wave packet
\begin{equation}
\Psi(x,y,t + \Delta t) = e^{-\frac{i}{\hbar}H\Delta t}\Psi(x,y,t)
\end{equation}
where the Hamiltonian $H$, assumed to be time-independent, is the one for low energy electrons in graphene
\begin{equation}
H = v_F \left(\vec{\sigma} \cdot\vec{p}\right) + V(x,y)\ {\bf I},
 \label{eqnum18}
\end{equation}
where $\vec{\sigma}$ is the usual Pauli vector, ${\bf I}$ is the
$2\times 2$ identity matrix and the wave functions are written as
pseudo-spinors $\Psi = (\Psi_A, \Psi_B )^T$, where $\Psi_A$
($\Psi_B$) is the probability of finding the electron in the
sub-lattice $A$ ($B$) of graphene. We separate the potential and kinetic energy terms of the time-evolution operator through the split-operator technique~\cite{degani, suzuki, chaves2015},

\begin{widetext}
\begin{equation}
\label{eq.SplitOperator}
\exp\left[-\frac{i}{\hbar}H\Delta t\right] = \exp{\left[-\frac{i}{2\hbar}V(x,y){\bf I}\Delta t\right]}\exp{\left[-\frac{i}{\hbar}v_F \vec{p}\cdot \vec{\sigma}\Delta t\right]}\exp{\left[-\frac{i}{2\hbar}V(x,y){\bf I}\Delta t\right]},
\end{equation}
\end{widetext}
where terms of order higher than $O(\Delta t^3)$ are neglected as an approximation. The advantage of this separation is that it allows us to perform multiplications in real and reciprocal space separately, thereby avoiding the need to write the momentum operator as a derivative by simply applying a Fourier transform to the functions and using the relation $\vec{p} = \hbar \vec{k}$. Furthermore, the exponential terms involving Pauli matrices can be re-written exactly as matrices~\cite{khamdam}, which further simplifies the calculations. We perform wave packet propagation with a time step as small as $\Delta t = 0.1$ fs and track the probabilities of finding the electron before and after the scattering region. The latter probability can also be interpreted as the transmission probability ($P$) through the scattering region. The criteria used here to define the space grid and time step is the same as employed in Refs. \cite{chaves, chaves2015, pereira, chaves2015energy}. In fact, a comparison between the results obtained by this method and parameters and those obtained by analytical solutions and tight-binding models in graphene-based systems has been recently shown in Ref. \cite{lavor}, where good agreement between results supports the validity of the methods and parameters used in the present work as well.

The obtained transmission probabilities $P$, as defined through wave packet propagation, are directly linked to the conductivity of the system. In fact, within the Landauer-B\"{u}ttiker formalism, the conductivity at a voltage $V$ and zero temperature is proportional to the transmission probability of a plane wave with energy exactly at $E = eV$. For non-zero temperature $T$, there exists an energy window of width approximately $k_B T$, within which the transmission probabilities of plane waves are integrated over energy to compute the conductance \cite{kwant}. In our method, however, we do not use a single plane wave but rather a wave packet with an average energy $E$, whose spatial width $a_x$ results in a distribution of energies around $E$. Since wave packets can be viewed as superpositions of plane waves, the wave packet propagation method employed here (as in Refs. \cite{kite, kramer, cavalcante}) effectively calculates the transmission probabilities of a combination of plane waves. Consequently, the energy window $k_B T$ in the Landauer approach is analogous to the energy spread of the wave packet in our approach. As a result, the conclusions drawn from the calculated $P$ at a given energy are directly related to the system's conductivity at a voltage where $eV$ is close to the average energy of the wave packet.

Notice that performing a Fourier transform while solving Eq. (\ref{eq.SplitOperator}) necessarily imposes periodic boundary conditions on our system \cite{chaves2015}. To avoid spurious reductions in transmission probability (or enhancements in reflection probability), we consider a very long grid in the propagation direction, with dimensions of 128 nm $\times$ 1024 nm, while the scattering centers are confined within an area of $A = 128$ nm $\times$ 40 nm centered around the $x = 0$ axis. Since periodic boundary conditions are naturally implied in both lateral directions, we have carefully verified that the propagated wave front behaves as expected for a graphene monolayer under such conditions: (i) in the absence of scattering centers, it propagates with a constant velocity $v_F$ and maintains its Gaussian shape and width throughout the entire time evolution, owing to the dispersionless nature of Dirac-like quasiparticles, and (ii) it returns to the system from the left boundary of the computational box (sketched in Fig. \ref{fig1}) once it reaches the right boundary. Nevertheless, periodicity in the propagation direction does not affect our calculations, as the computational box is sufficiently long to prevent the wave packet from reaching the right boundary before transmission probabilities are computed. All simulations were performed at a wave packet energy of $E = 100$ meV to ensure that the method remains within its valid range for the low-energy dispersion regime of graphene.

It is also worth pointing out that, in calculations of average values for propagated wave packets in systems such as monolayer graphene -- which is typically clean -- quantum fluctuations may play a role. This issue is thoroughly discussed, for example, in Refs. \cite{ishikawa2024overlap, ishikawa2024potential}. Nevertheless, transport property calculations, whether based on wave packet methods or the Landauer-B\"{u}ttiker formalism, have consistently been performed in recent years without accounting for quantum fluctuations \cite{rusin1, maks, kite, kwant, cavalcante, kramer}. The results and predictions obtained from these approaches have shown good agreement with experimental observations. Therefore, the effect of quantum fluctuations on the transport properties of the graphene system considered here is expected to be negligible. A deeper investigation of these effects on wave packet dynamics in monolayer graphene -- and how they could be experimentally detected -- is left as an exciting direction for future research.

\section{Results and discussions}

\begin{figure}[t]
\centering
\includegraphics[width=\linewidth]{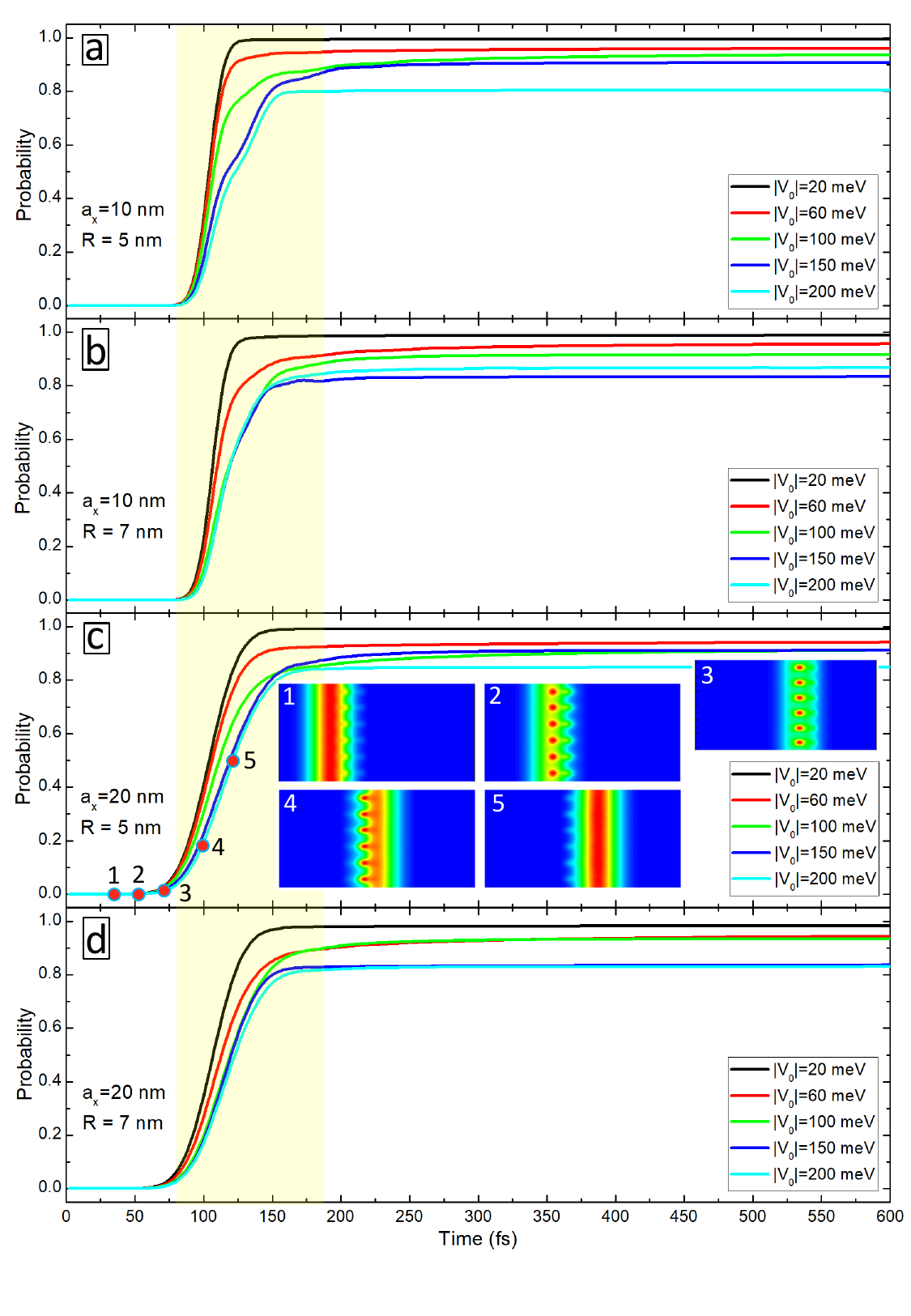}
\vspace{-1cm}
\caption{\label{fig2} Transmission probabilities of the wave packet as a function of time for Sample 1, for different values of the barrier height $|V_0|$. The wave packet width is $a_x=10$ nm (a, b) and $a_x=20$ nm (c,d), while the dot radius is $R=5$ nm (a,c) and $R=7$ nm (b,d). The shaded areas highlight the transmission probabilities when the wave packet is within the defected region. Insets in (c) shows the snapshots of the wave-packet propagation for parameters $a_x=20$ nm and $R=5$ nm at times indicated on the probability curve for $|V_0|=200$ meV. }
\end{figure}

As a reference for our further studies, we first investigate the case where the potential barriers are periodically arranged in the direction perpendicular to the wave packet propagation, without any spatial shift ($\delta = 0$; sample 1, see Fig. \ref{fig1}(a)). Figure \ref{fig2} presents the transmission probabilities $P$ of the wave packet as a function of time for different values of the barrier height $|V_0|$ in this configuration. For smaller dot radii, the transmission probability decreases with increasing barrier height (see Fig. \ref{fig2}(a)). As the size of the barrier increases, a non-monotonic dependence of $P$ on $|V_0|$ is observed (Fig. \ref{fig2}(b)). For example, the transmission probability for $|V_0| = 200$ meV is higher than that for $|V_0| = 150$ meV in Fig. \ref{fig2}(b). The effect of the barriers on the wave packet dynamics becomes more pronounced for wider wave packets, as shown in Figs. \ref{fig2}(c,d). For this particular wave packet width, a non-monotonic dependence of $P$ on the barrier height is observed even for the smallest considered dot size (Fig. \ref{fig2}(c)). Panels 1-5 in the inset of Fig. \ref{fig2}(c) illustrate the wave packet's propagation dynamics through the barrier region. Although the transmission probability remains high ($P > 80$\%) for the chosen parameters, significant distortion of the wave packet occurs within the defect area.

\begin{figure}[b]
\centering
\includegraphics[width=\linewidth]{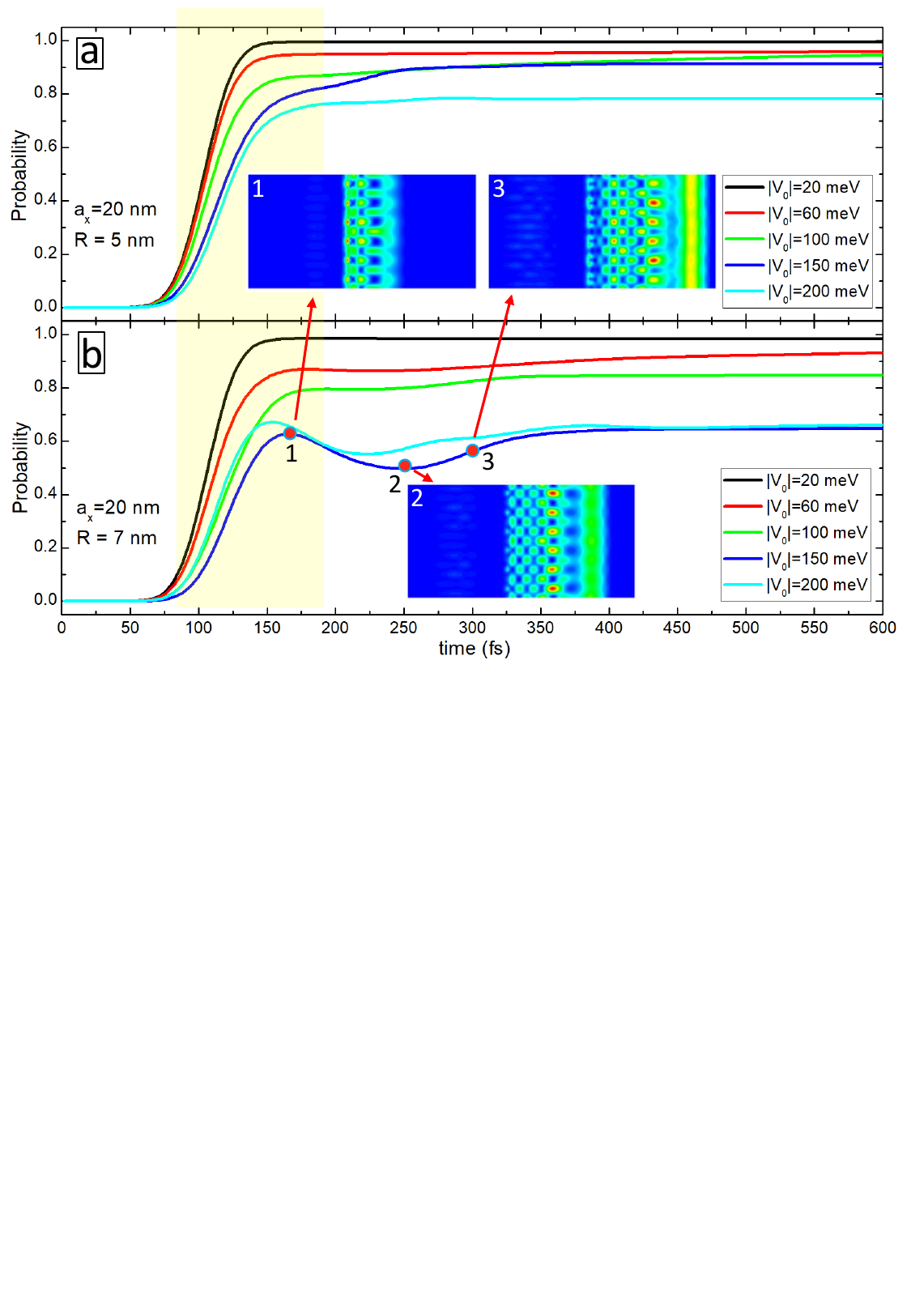}
\vspace{-7cm}
\caption{\label{fig3} Probabilities of finding the electron after the barrier region as a function of time for Sample 2, for different values of the barrier height $|V_0|$. Insets shows the snapshots of the wave-packet propagation for parameters $a_x=20$ nm and $R=7$ nm at times indicated on the probability curve for $|V_0|=150$ meV.}
\end{figure}

Figure \ref{fig3} shows the transmission probabilities of the wave packet in sample 2, where dots of the same potential polarity are shifted by a distance of $\delta = 6R$ relative to the dots of opposite polarity (see Fig. \ref{fig1}(b)). Two main differences are observed when comparing these results with those of sample 1. First, the transmission probabilities of the wave packet decrease across all considered values of $V_0$ in sample 2. This suggests a higher degree of wave packet scattering or increased influence of the potential barriers in this configuration. Second, additional features appear on the probability curves, indicating more complex interactions within the sample that affect the wave packet's behavior. This is illustrated in panels 1-3 of the inset in Fig. \ref{fig3}, where snapshots of the wave packet distribution are shown at times corresponding to the features observed on the probability curve. For example, the minimum observed on the probability curve (point 2) corresponds to the wave packet scattering from the second row of dots. A significant portion of the wave packet is reflected by the defected area, thereby reducing the overall transmission. Thus, the dynamics of the wave packet are significantly influenced by the simple act of shifting dots with the same potential polarity, as evidenced by decreased transmission probabilities, increased scattering, and the emergence of complex features in the probability curves.

\begin{figure}[t]
\centering
\includegraphics[width=\linewidth]{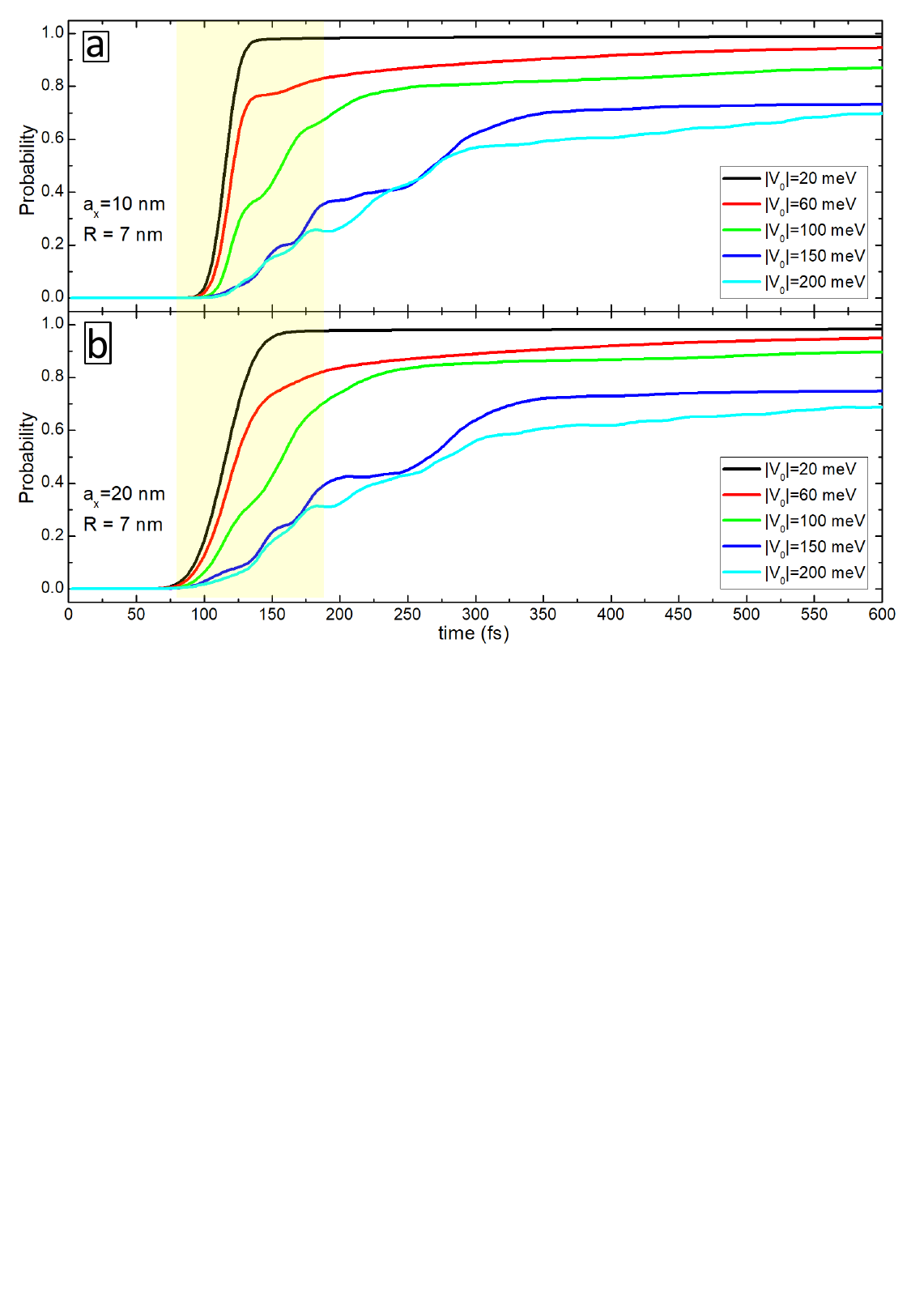}
\vspace{-7cm}
\caption{\label{fig4} Transmission probabilities of the wave packet as a function of time for Sample 3 for different values of the barrier height $|V_0|$. The radius of the dots is $R=7$ nm and the width of the wave packet is $a_x=10$ nm (a) and $a_x=20$ nm (b).}
\end{figure}

The transmission probability decreases further as the number of scattering centers increases, due to the prolonged interaction between the wave packet and the defects. This is illustrated in Fig. \ref{fig4} for sample 3, which includes an additional row of mutually shifted dots (see Fig. \ref{fig1}(c)). The reduction in transmission is particularly noticeable when the wave packet is within the defected region, where the defects disrupt its propagation. This reduction arises from scattering and interference caused by the structural irregularities. As the barrier height increases, the transmission probabilities decrease further, consistent with previous observations. A higher barrier height leads to increased reflection and scattering of the wave packet, making it more difficult for the packet to tunnel through the defected region and thereby reducing the overall transmission.

\begin{figure}[b]
\centering
\includegraphics[width=\linewidth]{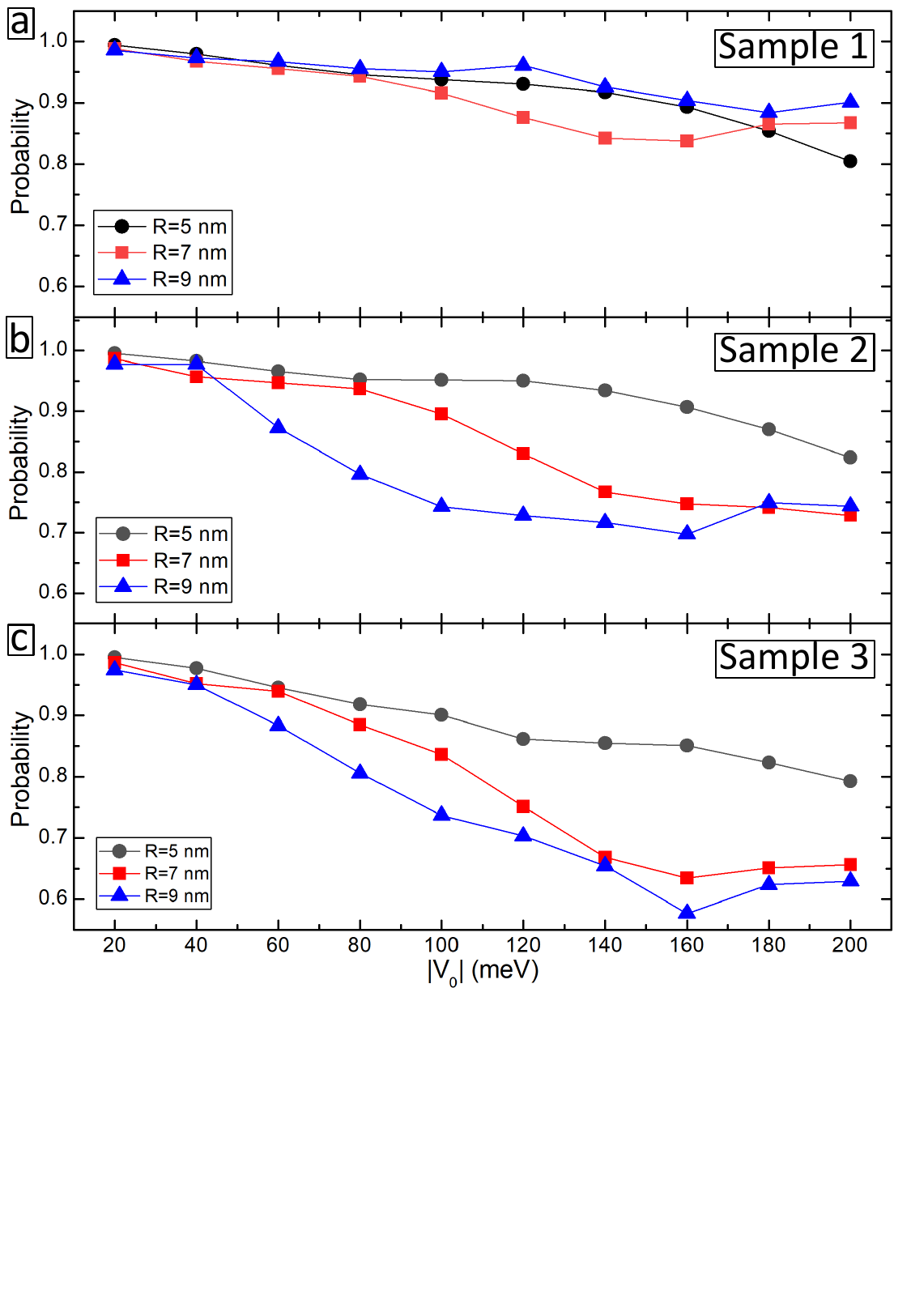}
\vspace{-3.5cm}
\caption{\label{fig5} Transmission probabilities of the wave packet in sample 1 (a), 2 (b) and 3 (c) as a function of barrier height $|V_0|$ for three different values of the dot radius. The width of the wave packet is set to $a_x=10$ nm.}
\end{figure}

Figure \ref{fig5} summarizes our findings on how the rearrangement of potential defects affects wave packet dynamics in graphene. The figure presents the transmission probability of the wave packet as a function of barrier height $V_0$ for samples 1-3, illustrating the influence of different defect configurations on transmission behavior. The most apparent effect of defect rearrangement is a reduction in the wave packet's transmission. For instance, the transmission probability at $V_0 = 200$ meV decreases from 0.8 in sample 1 to 0.6 in sample 3, highlighting the impact of defect configuration on the wave packet's ability to transmit. Secondly, the effect of dot size becomes more pronounced as the spacing between the dots increases. For example, in sample 1, the maximum difference in transmission probability between dot radii of $R = 5$ nm and $R = 7$ nm is 0.08, whereas in sample 3, the same difference increases to 0.22. In configurations where the dots are more sparsely distributed, the wave packet interacts with each defect individually for a longer duration, amplifying the influence of dot size on transmission. Larger dots create stronger scattering potentials, which significantly disrupt the wave packet's trajectory and reduce transmission. In contrast, when the dots are closely spaced, the individual effect of each dot is less distinct, and the overall transmission behavior is more strongly influenced by the collective scattering from the entire defected region.

\begin{figure}[t]
\centering
\includegraphics[width=\linewidth]{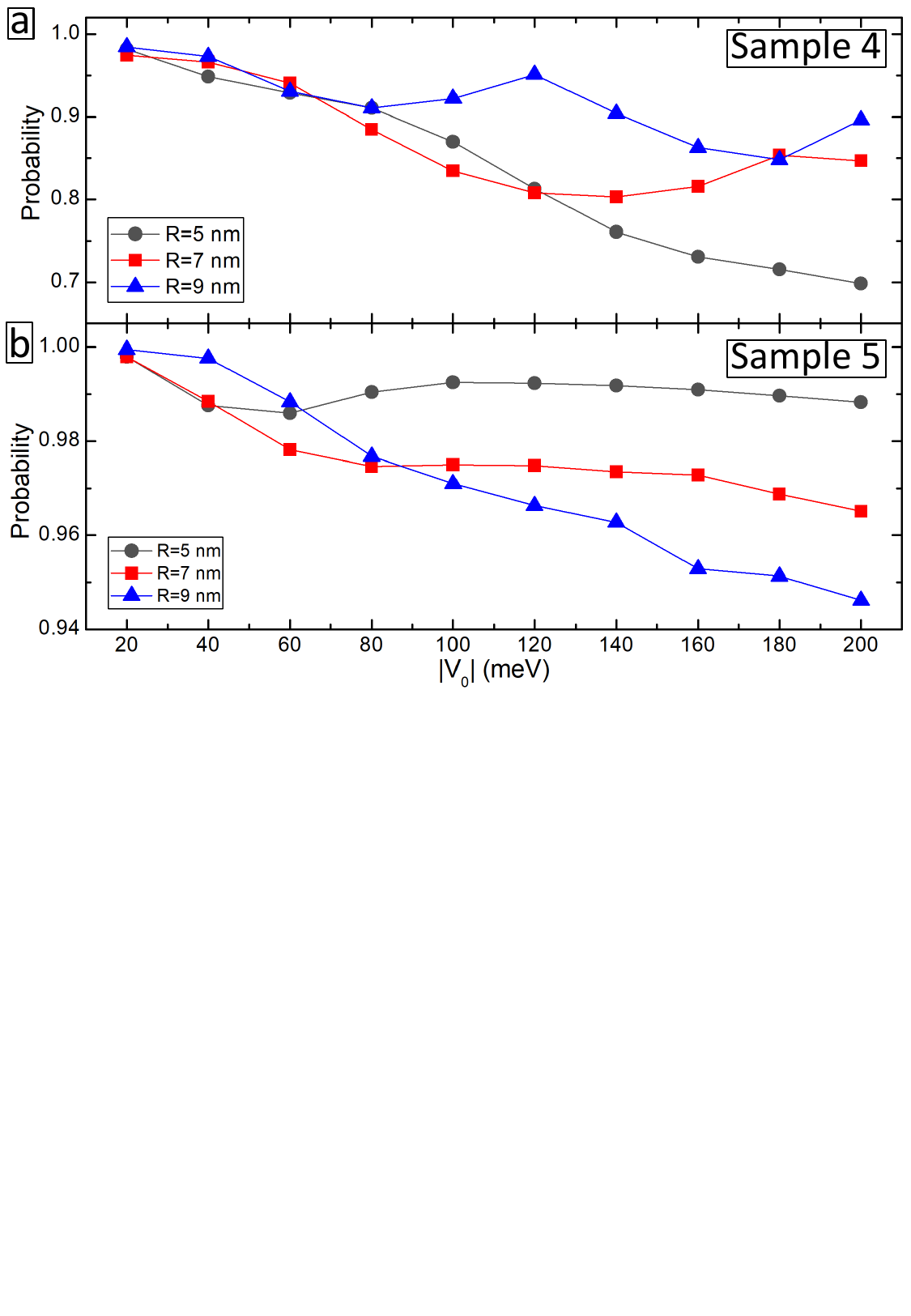}
\vspace{-6.5cm}
\caption{\label{fig6} Transmission probabilities of the wave packet in sample 4 (a) and 5 (b) as a function of energy for three different values of the dot radius. The width of the wave packet is set to $a_x=10$ nm.}
\end{figure}

Finally, we explore the impact of potential polarity alterations on the dynamics of the wave packet, as demonstrated in samples 4 and 5. Sample 4 features a symmetric vertical arrangement of alternating potentials, whereas sample 5 presents a staggered, zigzag pattern, introducing asymmetry in the wave packet's interaction with the potential barriers. Figure \ref{fig6} presents the transmission probabilities of the wave packet as a function of barrier height $V_0$ for these samples, for three different dot sizes. This comparison highlights how barrier height and dot size influence transmission behavior in each configuration. The variation in arrangement leads to distinct transmission and scattering responses between the two samples. Interestingly, for sample 4, the transmission probability increases with the radius of the dots at higher values of $V_0$ (see Fig. \ref{fig6}(a)). In contrast, for sample 5, an increase in dot size results in reduced transmission probabilities. This opposite trend underscores the critical role of dot arrangement in determining how the wave packet interacts with the potential barriers in each sample.

The observed transmission behavior can be attributed to a combination of interference effects and quasi-resonant tunneling mechanisms. The non-monotonic dependence of the transmission probability on barrier height and dot size arises from multiple scattering events within the periodic potential landscape. As the wave packet propagates through the array of potential barriers, partial constructive or destructive interference emerges depending on the energy and spatial configuration, leading to enhancement or suppression of transmission. Moreover, due to the periodicity of the barriers and the finite width of the wave packet (which corresponds to an energy spread), quasi-resonant tunneling can occur at specific energies, manifesting as localized peaks in the transmission probability. These effects highlight the wave nature of charge carriers in graphene and underscore the importance of barrier geometry and symmetry in modulating electronic transport.

\section{Conclusions}

In this work, we used the Dirac continuum model to study the propagation of wave packets representing low-energy electrons in monolayer graphene with periodically arranged potential barriers. The system studied here provides insights into the effects of spatially varying electrostatic potentials originating from positive and negative charge puddles commonly found in real graphene samples. Our simulations reveal that both the size and arrangement of scattering centers significantly influence transmission probabilities. Increasing the barrier height leads to non-monotonic changes in transmission, with larger wave packets being more strongly affected. The configuration of potential barriers -- whether symmetric or staggered -- also plays a crucial role, with staggered arrangements leading to greater transmission suppression. These findings deepen our understanding of wave packet dynamics in graphene and offer useful guidance for the design of future graphene-based electronic devices.

\section{Acknowledgements}

AC acknowledges financial support from the Brazilian National Council for Scientific and Technological Development (CNPq), through the UNIVERSAL and PQ programs, and from the Brazilian Coordination for the Improvement of Higher Education Personnel (CAPES), through the PROBRAL program.

\end{document}